\documentclass{appolb}
\usepackage{exscale}                  
\usepackage[intlimits]{amsmath}       
\usepackage{amsfonts}
\usepackage{amssymb,amscd}
\usepackage[dvips]{epsfig}                   
\usepackage{array}
\usepackage{wrapfig}
\usepackage{bbm}
\usepackage{multirow}
\usepackage{cite}

\newcommand{\beqa}{\begin{eqnarray}}
\newcommand{\eeqa}{\end{eqnarray}}
\newcommand{\beq}{\begin{equation}}
\newcommand{\eeq}{\end{equation}}

\def\eq#1{(\ref{#1})}

\begin{document}

\title{Baby steps beyond rainbow-ladder%
\thanks{Presented by R. Williams at ``Excited QCD'' 2009, February 8--14, Zakopane, Poland}%
}
\author{Richard Williams
\address{Institute for Nuclear Physics, 
 Darmstadt University of Technology, 
 Schlossgartenstra{\ss}e 9, 64289 Darmstadt, Germany}
\and
Christian~S.~Fischer
\address{Institute for Nuclear Physics, 
 Darmstadt University of Technology, 
 Schlossgartenstra{\ss}e 9, 64289 Darmstadt, Germany}
\address{GSI Helmholtzzentrum f\"ur Schwerionenforschung GmbH,\\ 
  Planckstr. 1  D-64291 Darmstadt, Germany.}
}
\maketitle

\begin{abstract}
We discuss the impact of including corrections beyond single gluon
exchange in light mesons within the nonperturbative
framework of Dyson-Schwinger equations (DSE) and Bethe-Salpeter
equations (BSE). We do this by considering unquenching effects in the form
of hadronic resonance contributions, notably pion exchange, and by the
inclusion of the dominant gluon self-interactions to the quark-gluon vertex.  
Thus we make steps towards an \emph{ab initio}
description of light mesons by functional methods.
\end{abstract}


\section{Introduction}
With the absence of quarks and gluons in the physical spectrum, we are
forced to probe their interaction at low energies by the study of colourless composites 
of these particles. This information is encoded in the Green's
functions of our QFT, in particular the four quark-scattering matrix and
the
quark-gluon vertex. Faced with the richness of the hadronic spectrum, it is not
unreasonable to draw a comparison with the complex nonperturbative
particulars of these Green's functions. In this talk we will consider
the simplest bound states -- the light mesons -- as our probes
and investigate how their properties depend on information 
present in the quark-gluon vertex.

\begin{figure}[b]
  \begin{center}
\begin{eqnarray}
  \nonumber\\[-9mm]
  \includegraphics[width=0.66\columnwidth]{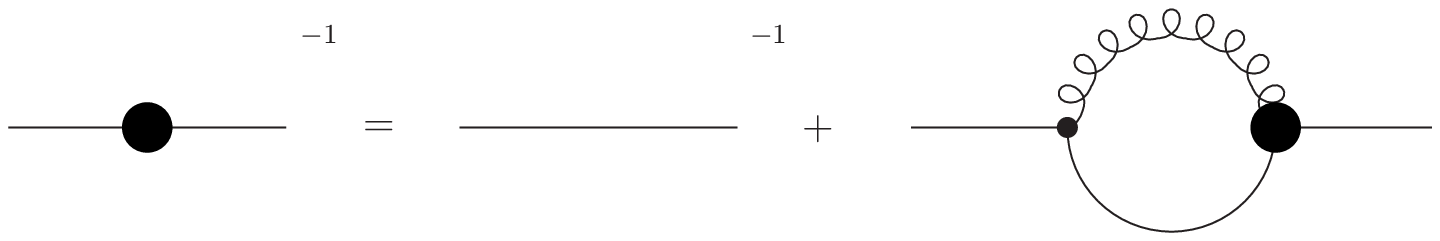}\label{fig:quarkdse}\nonumber\\[-13mm]\\[8mm]
\includegraphics[width=0.71\columnwidth]{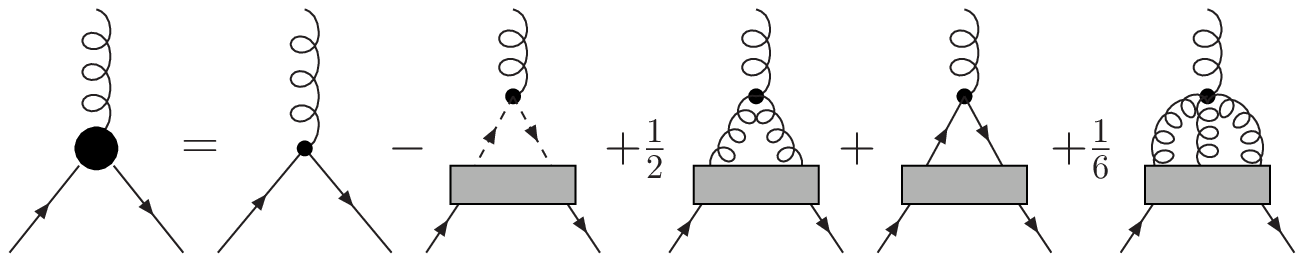}\label{dse1}\nonumber\\[-13mm]\\[8mm]
\includegraphics[width=0.71\columnwidth]{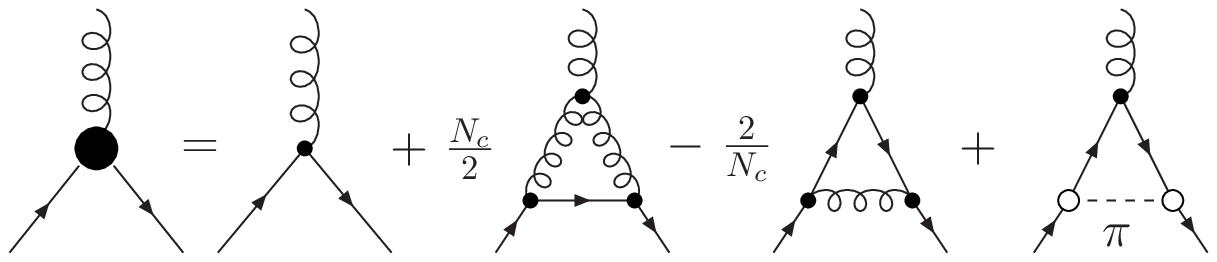}\label{dse2}\nonumber\\[-13mm]\\
\nonumber
\end{eqnarray}
\end{center}
\caption{
DSEs for: (\ref{fig:quarkdse}) fully dressed quark propagator;
(\ref{dse1}) full quark-gluon vertex;
(\ref{dse2}) truncated quark-gluon vertex. Internal
propagators are dressed, with gluons shown by wiggly lines, quarks by
straight lines and dashed lines mesons. White-filled circles show
meson amplitudes whilst black-filled represent vertex dressings.
\label{fig:qgdse}}
\end{figure}
\section{Dyson-Schwinger equations}
To solve the DSEs, shown for the inverse quark propagator and quark-gluon vertex
by (\ref{fig:quarkdse}) and (\ref{dse1}) of Fig.~\ref{fig:qgdse}, we need to introduce a truncation at some point
in the infinite tower. This need for a truncation also applies to the BSE
since therein we must specify the (2PI) four-point scattering kernel.  This is a
delicate process since we must be careful to
preserve various symmetries of the theory. The most important of
these relates to chiral symmetry, expressed via the axial-vector
Ward-Takahashi identity (axWTI). This underpins the observed
mass spectrum of the light mesons,  and ensures that pions are indeed
the (pseudo)-Goldstone bosons of the theory.
This identification is a necessary feature of any serious model 
and must be exhibited by any truncation of the BSEs and DSEs.

The
simplest truncation that satisfies this criterion is that of
Rainbow-Ladder (RL) whereby the full quark-gluon vertex is replaced by a
bare vertex, see \emph{e.g.}~\cite{Maris:1997tm}. The axWTI preserving 
kernel in the BSE then corresponds to a single gluon exchange,
re-summed to all orders thus providing the `ladder'. These 
RL models are designed to reproduce predominantly s-wave mesons due to the simple vector-vector structure of the
interaction. To compensate for this simplicity one constructs a
phenomenological model for the gluon, effectively subsuming 
additional vertex corrections from the Yang-Mills sector and unquenching effects. 

To separate the phenomenological from the \emph{ab initio}, we must investigate
the quark-gluon vertex and determine the impact of corrections beyond
tree-level to our quarks and mesons. Such studies have been made
\emph{e.g.}~\cite{Bender:1996bb,Bender:2002as,Bhagwat:2004hn,Watson:2004kd,Bhagwat:2004kj,Matevosyan:2006bk,Chang:2009zb}.
Following the analysis of~\cite{Alkofer:2008tt,Fischer:2007ze} we approximate the full
DSE \eq{dse1} with the (nonperturbative) one-loop structure of \eq{dse2}. Here the first
`non-Abelian' loop-diagram in \eq{dse2} subsumes the first two diagrams in the full
DSE to first order in a skeleton expansion of the four-point functions.
We neglect the two-loop diagram in the full DSE \eq{dse1}, which is
justified for small and large momenta~\cite{Alkofer:2008tt,Bhagwat:2004kj}. 
The remaining `Abelian' contributions are split
into the non-resonant second loop-diagram in \eq{dse2} and a third
diagram containing effects due to hadron back-reactions.

In the next section we consider these resonance contributions, using a RL truncation for the non-resonant
parts. We follow this by exploring a new truncation in which leading 
non-resonant parts from the non-Abelian vertex are included self-consistently in the quark DSE and meson BSE.  

\section{Including unquenching effects}\label{sec2}
The prescription for including pion degrees of freedom in the DSEs and
BSEs in a manner consistent with the axWTI have been proposed and investigated
in~\cite{Fischer:2007ze,Fischer:2008sp,Fischer:2008wy}, with the resultant system of equations depicted
by
\begin{eqnarray}
  \includegraphics*[width=0.80\columnwidth]{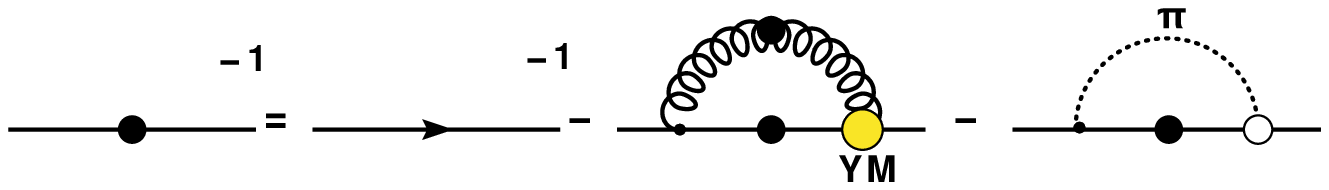}\nonumber\\
  \includegraphics*[width=0.75\columnwidth]{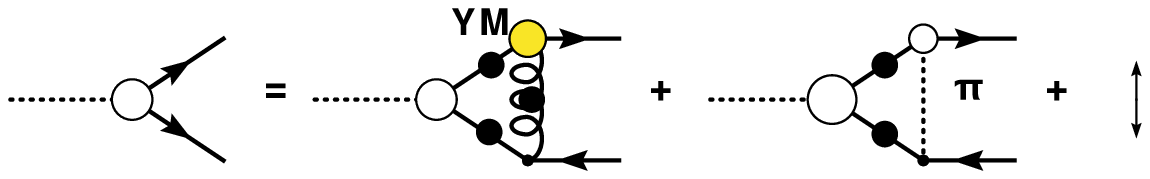}\nonumber
\end{eqnarray}
The first loop diagram of both 
equations relates to the usual rainbow-ladder, where the infrared
suppressed gluon is
enhanced by the vertex dressing, indicated by \emph{YM}.
The second diagram contributing to the quark DSE and meson BSE
represents the back-reaction of the pion onto the quark. This requires input of the
quark-pion vertex, which we parameterise by a chiral approximation of
the leading pion Bethe-Salpeter amplitude~\cite{Fischer:2008wy}
\begin{equation}
\Gamma^j_{\pi}(p;P) = \tau^j \gamma_5 B_\chi(p^2)/f_\pi\;. \label{piapprox}
\end{equation}
Here $B_\chi(p^2)$ is the scalar dressing function of the quark propagator
in the chiral limit. The effects of neglecting the three sub-leading 
amplitudes have been quantified for a real-value approximation in 
Ref.~\cite{Fischer:2007ze}, and found to be on the level of a few
percent.
The advantage of the approximation in (\ref{piapprox}) is that we can
then directly calculate the quark propagator in the complex plane. 

We need to specify the gluon propagator and a quark-gluon
vertex that subsumes the non-resonant parts of the interaction into an
effective rainbow-ladder model. Since here we wish to employ a gluon
propagator as calculated from its DSE, we employ the soft divergent
model (SD) for the quark-gluon vertex as described
in~\cite{Alkofer:2008tt,Fischer:2008wy}.  

We calculated a range of meson observables and observe that the effect of the pion
back-reaction has only a small impact on the pion mass itself.
The impact of including pion-cloud effects on the leptonic
decay constant is more substantial, with effects of the order of
$10$~MeV:

\begin{center}
\begin{tabular}{@{}c|cc|cc|ccc}
  Model & $M_\pi$ & $f_\pi$ & $M_\rho$ & $f_\rho$ &$M_\sigma$ &
   $M_{a_1}$ & $M_{b_1}$\\
\hline
\hline
 quenched  & 125        & 102 & 795 & 159& 638  & 941 & 879\\
 unquenched& 138$^\dag$ & 93.8$^\dag$& 703 & 162& 485  & 873 & 806 \\
\hline
\hline
PDG~\cite{Allton:2005fb}& 138 & 92.4& 776 & 156&400--1200 &1230 &1230 \\
\end{tabular}
\end{center}

For the remaining heavier mesons, the common trend is that the inclusion
of the pion cloud decreases the masses by $100$--$200$~MeV. 
Most notable of these are for the rho, where we see
that unquenching from the pion-cloud yields a bound-state that is $\sim90$~MeV
lighter. This will prove important in what follows when we consider gluon
self-interaction contributions.

It is clear, however, that in order to reproduce the rich spectrum of
light mesons we need to include spin dependent contributions 
from the Yang-Mills part of the quark-gluon vertex. We now consider this
in the absence of pion-cloud effects in the next section. 

\section{Including gluon self-interactions}\label{sec3}
Looking back to our proposed truncation of the quark-gluon vertex, we
identified the second loop diagram of (\ref{dse2}) as the dominant
contribution. Ignoring all other contributions the resulting equation,
coupled with the quark DSE, may be solved numerically provided we know
the gluon propagator and three-gluon vertex dressing. 
Since we work in Euclidean space we need the 
quark propagator for complex values of its momenta. This
means our quark-gluon vertex must also be solved for complex  momenta. Through judicious choice of momenta in both the quark and vertex DSE, this can be performed without unconstrained analytic
continuation of the gluon propagator and three-gluon vertex. 
The axWTI preserving truncation for the BSE\cite{Bender:1996bb,Munczek:1994zz},
consistent with our choice of the quark-gluon vertex is~\cite{Maris:2005tt} 
%
\begin{eqnarray}
	\begin{array}{c}
	\includegraphics[scale=0.6]{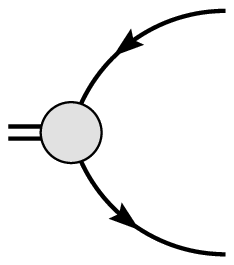}\\
	\end{array}
	&=& 
	\begin{array}{c}
	\includegraphics[scale=0.6]{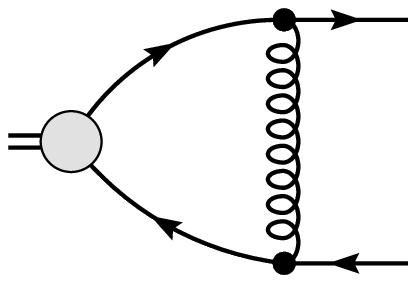}\\
	\end{array}
	+
	\begin{array}{c}
	\includegraphics[scale=0.6]{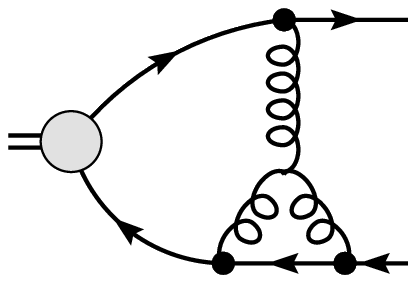}\\
	\end{array}
	+
	\begin{array}{c}
	\includegraphics[scale=0.6]{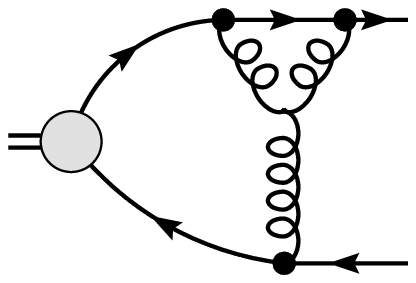}\\
	\end{array}
	\nonumber
	\end{eqnarray}

Since this
work is a preliminary study of a new and sophisticated
truncation~\cite{Fischer:2009jm} we choose to simplify things further by employing 
a simple momentum dependent ansatz for the gluon~\cite{Alkofer:2002bp}
\begin{equation}
  \alpha Z(q^2) =\left(\pi D /\omega^2\right)\;q^4\;  e^{-q^2/\omega^2}\;,\label{eqn:gluon}
\end{equation}
with two parameters $D$ and $\omega$ which provide for the scale and 
strength of the effective gluon interaction. Naturally, this ansatz
provides only a first step towards a full calculation including input from the 
DSEs for the three-gluon vertex and the gluon propagator. Nevertheless
we believe that the ansatz \eq{eqn:gluon} is sufficient to provide for reliable 
qualitative results as concerns the effects due to the non-Abelian diagram onto 
meson properties. 

The results of our calculation follow,
wherein we compare our truncation
including gluon self-interaction effects in all twelve tensor 
structures of the quark-gluon vertex to that of RL with only vector-vector interactions. 
\begin{center}
\begin{tabular}{@{}c||cc|cc|cc|ccc}
  Model & $\omega$ & $D$ & $m_\pi$ & $f_\pi$  &
$m_\rho$  & $f_\rho$  & $m_\sigma$ & $m_{a_1}$ &
$m_{b_1}$\\
  \hline\hline
R-L& \multirow{2}{*}{$0.50$} & \multirow{2}{*}{$16$} & $138$  & $94$ 
&  $758$  & $154$ &  $645$ & $926$ & $912$\\
  BTR   &        &      & $138$  & $111$   &  $881$  & $176$
&   $884$ & $1055$ & $972$\\
  \hline
  R-L   & \multirow{2}{*}{$0.45$} & \multirow{2}{*}{$25$} & $136$ &
$92$      &  $746$  & $149$   & $675$ & $917$ & $858$\\
  BTR   &        &      & $142$ & $110$  &  $873$  & $173$
&  $796$ & $1006$ & $902$\\
  \hline\hline
PDG~\cite{Allton:2005fb} &     &      & $138$  & $92.4$    & $776$ & $156$  & $400$--$1200$ & $1230$ & $1230$ \\
\end{tabular}
\end{center}
The model parameters 
$\omega$ and $D$ were tuned such that for the latter we obtain 
reasonable pion observables, and the quark mass is fixed at $5$~MeV.

What we find is that the mass of the $\rho$-meson 
is enhanced by $\sim120$~MeV as compared to pure RL. This is intriguing
since it has long been suspected that corrections beyond 
RL approximately cancel in the
$\rho$~\cite{Bender:1996bb,Bender:2002as,Eichmann:2008ae}.
This is supported by known estimates of mass shifts from the resonant
and non-resonant Abelian diagrams in \eq{dse2}, calculated at
$\sim90$~MeV (see previous section) and $30$~MeV~\cite{Watson:2004kd}
respectively

\section{Conclusions} We presented a study of light mesons using two
truncations of the Bethe-Salpeter equations beyond RL. These considered
unquenching effects associated with the pion cloud, and the dominant
non-Abelian corrections to the quark-gluon vertex stemming from gluon
self-interactions. For the latter truncation, we obtain masses for the rho
meson of $\sim900$~MeV, consistent with quenched lattice simulations. The
subsequent inclusion of unquenching effects and other non-resonant
contributions to the quark-gluon interaction brings the rho mass back
to its physical value thus supporting a long suspected 
cancellation mechanism.  An investigation based on Yang-Mills DSE results
together with unquenching effects is currently underway.

\section{Acknowledgements}
This work was supported by the Helmholtz-University Young Investigator
Grant number VH-NG-332 and by the Helmholtz International Center for FAIR
within the framework of the LOEWE program (Landesoffensive zur Entwicklung
Wissenschaftlich-\"Okonomischer Exzellenz) launched by the State of Hesse.

\end{document}